# Deep learning for visualization and novelty detection in large X-ray diffraction datasets

Lars Banko[1], Phillip M. Maffettone[2], Dennis Naujoks[3], Daniel Olds[2], Alfred Ludwig[1,3]

1. Materials Discovery and Interfaces, Institute for Materials, Ruhr-University Bochum, 44801 Bochum, Germany
2. National Synchrotron Light Source II, Brookhaven National Laboratory, Upton, New York 11973, USA
3. Center for Interface-Dominated High Performance Materials, Ruhr-University Bochum, 44801 Bochum, Germany

**Abstract**

We apply variational autoencoders (VAE) to X-ray diffraction (XRD) data analysis on both simulated and experimental thin-film data. We show that crystal structure representations learned by a VAE reveal latent information, such as the structural similarity of textured diffraction patterns. While other artificial intelligence (AI) agents are effective at classifying XRD data into known phases, a similarly conditioned VAE is uniquely effective at knowing what it doesn't know, rapidly identifying novel phases and mixtures. These capabilities demonstrate that a VAE is a valuable AI agent for materials discovery and understanding XRD measurements both 'on-the-fly' and during *post hoc* analysis.

# Introduction

Innovations in high-throughput and autonomous experimentation[1-4] are exceedingly increasing the acquisition rate of data, particularly in the case of XRD. Manual analysis of combinatorial datasets is a challenging cognitive task that requires the ability to recognize patterns under the awareness of several constraints[5]. Recent progress has been made in using AI for crystal structure classification[1-4], and integrating the former with autonomous experimentation[6].

The success of autonomous experimentation depends on the prior information (e.g. simulated X-ray diffractograms of expected phases from crystallographic database entries) that are used to develop an AI agent. These priors stem from expert knowledge and expectation, and as such, even well trained, accurate AI systems can come to false conclusions when encountering previously unseen data or novelty. Here, we define novelty as XRD patterns that are unknown to the AI, e.g. structures that were absent in the training data or XRD patterns of phase mixtures which were not considered in the training data. In the context of materials discovery, material novelty comprises unreported materials of certain chemical composition and crystal structure.

The recognition of novelty in large amounts of high-throughput data is a key challenge for scientists, and subsequently the tools that they use. An AI agent that is sensitive to novelty could inform scientists about experiments that mandate further investigation due to failure modes or material novelty. In this way, scientists can invest essential resources into the most promising areas and focus their manual effort where human experience is valued most.

Here, we use a variational autoencoder (VAE) (Figure 1a) trained on a synthetic dataset as a prior[7,8] to solve a set of common challenges in XRD analysis. Firstly, we identify regions of similarity and potentially degenerate solutions in our prior. Next, we develop a dynamic visualization tool for experimental XRD patterns and the VAE latent space. While this offers a correlation with structural classification, we show that visualizing the latent space with respect to the reconstruction error of the VAE allows for novelty detection during an experiment.

# Results and discussion

Autoencoders are a class of self-supervised artificial neural networks that compress input data into a lower dimensional feature-value space (latent space) and then decodes this latent representation back into its original space. These have been broadly applied for the purpose of dimensionality reduction, denoising[9], anomaly[10] and novelty detection[11] and specific materials science applications such as detecting phase transitions[12], and translating between different dimensional representations, i.e., 2-d images and 1-d spectra[13,14]. VAEs are a special class of autoencoders that approximate a posterior over latent random variables[15,16], jointly optimizing the reconstruction of the input and the Kullback-Leibler (KL) divergence between the latent representation and a smooth (often normal) distribution. This later loss function acts as a regularization mechanism, resulting in an efficient distribution over the latent space. The resulting probabilistic latent space has made these approaches popular for generating new data based on the prior training data[17], and inverse design in materials science[18,19].

We first study the behavior of a VAE on XRD data with regards to latent space distribution using a synthetic dataset of 15000 one-dimensional XRD patterns covering three spacegroups (Fm$\bar{3}$m (ISCD: 108308), Im$\bar{3}$m (ISCD: 108347), P6$_3$/mmc (ISCD: 622438), 5000 XRD patterns each). The dataset includes different aberrations that are frequently encountered in thin-film XRD patterns such as texture and preferred orientation, peak broadening, and peak shifting[8]. Figure 1c shows the latent space distribution of the validation dataset, color-coded by the spacegroup labels. The latent space representation provides direct visual evidence of the clustering properties of the encoder model and distribution of main reflection axes in the XRD patterns (Figure 1b). Classification boundaries are mapped over this latent space to provide a visualization of structural similarity (Figure 2). In the latent feature space, proximity is a first indicator of structure type. $P6_3/mmc$ and $Fm\bar{3}m$ show an overlapping region which stems from homometrics: similar patterns that arise from distinct structures with preferred orientation in the training dataset. The XRD patterns, marked by green, red and blue crosses in Figure 1c, are shown in Figure 1d and highlight the similarity of $Im\bar{3}m$ and $P6_3/mmc$ for texturing along (020) and (102), respectively. By examining the latent space distribution with respect to the location of maximal intensity in the XRD patterns, we can see that the latent space is organized according to the main reflection axes of the crystal structures (Figure 1b). This demonstrates how the learned latent representation can directly point towards possible structural ambiguities in the prior, and creates an explainable AI tool for understanding what determines location in latent space.

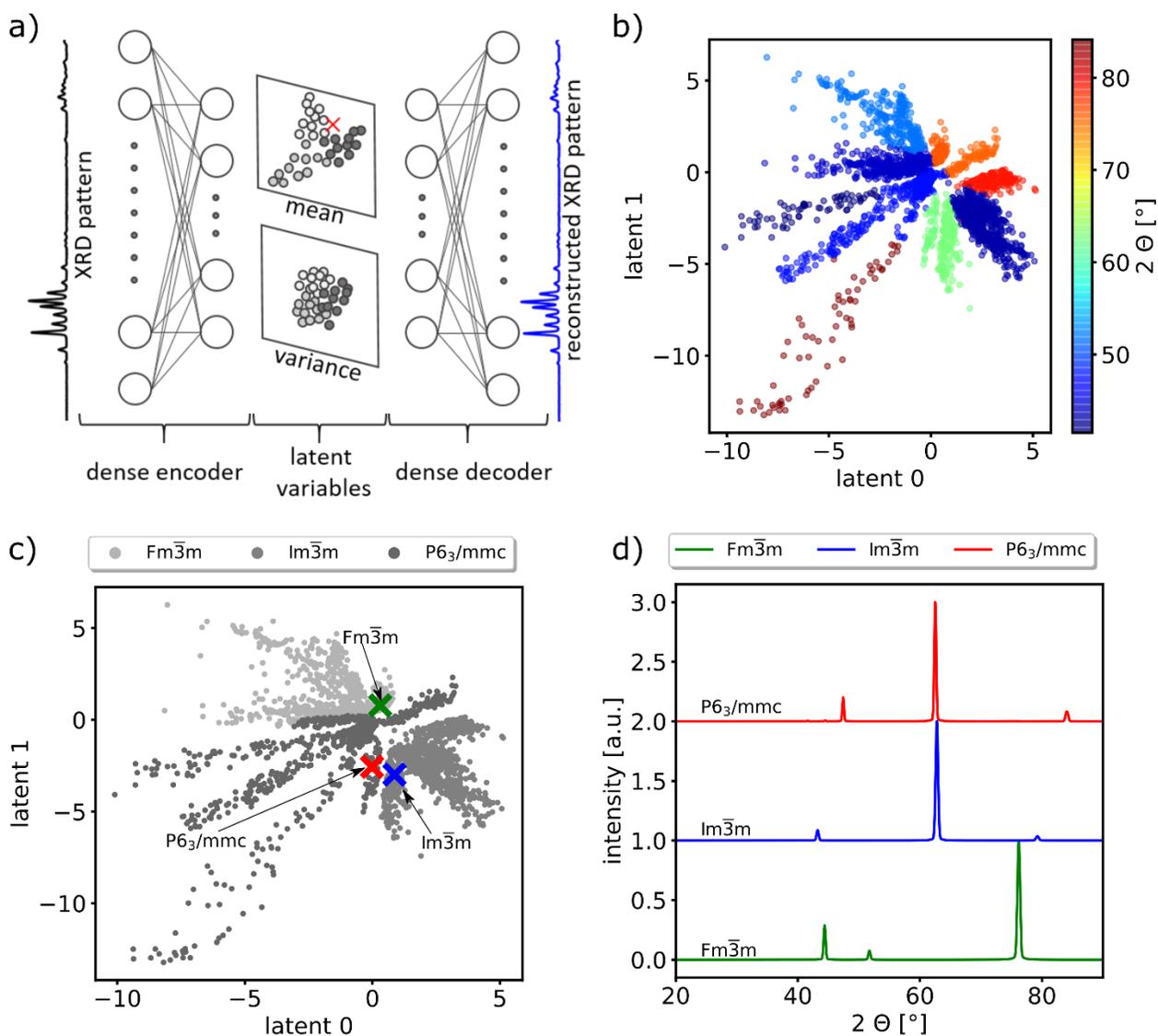

*Figure 1: a) Schematic VAE architecture. XRD patterns are encoded into a low dimensional representation (mean, variance). The red cross marker indicates the latent space position of the encoded XRD pattern with respect to the prior (circles). The latent vector is decoded into a reconstructed XRD pattern. b) Latent space embedding with color-coded diffraction angle 2Θ of maximum XRD intensity. c) Latent space embedding of an exemplary synthetic dataset. Im$\bar{3}$m is clearly separated from Fm$\bar{3}$m and P6$_3$/mmc. The x-markers show the latent space position of the corresponding XRD patterns in d). d) XRD patterns of the x-marked latent space positions in c). The latent space embedding in b) clearly shows the main reflection axes of the different crystal structures, i.e. P6$_3$/mmc has six main reflection axes in the angular range from 20 – 90 °2Θ (for Cu Kα). It further elucidates possible ambiguities between different structures that exhibit a preferred orientation: Fm$\bar{3}$m (111) and P6$_3$/mmc (002) as well as Im$\bar{3}$m (020) and P6$_3$/mmc (102) have peaks at a similar diffraction angle. This is important during the pattern matching task, as an experimental XRD pattern could be a result of either structure.*

By interfacing a VAE trained on synthetic data with an experimental datastream[20], this latent representation can visualize a measurement in real time. To demonstrate how this visualization behaves with new or unknown phases, a set of 1000 XRD patterns of a novel phase, unknown to the model, with $P4_2/mnm$ spacegroup (ICSD: 601378) are generated using the same parameter variation as in the training dataset. In Figure 2, the reconstruction error is mapped over the latent space for the test set of all phases (known and unknown/ novel). The decision boundaries of a KNN classifier, trained on the latent representation are outlined. While it could be naively assumed from the visualized location in latent space (Figure 2b) that the $P4_2/mnm$ phase was the $Im\bar{3}m$ or $P6_3/mmc$ phase, the reconstruction error increases by an order of magnitude, which means that the naive assumption is most likely wrong. This indicates a posterior distribution which does not capture the information contained in the input XRD pattern: this is new or novel to the model. The model encodes a latent representation for the $P4_2/mnm$ phase near the $P6_3/mmc$ and $Im\bar{3}m$ phases which suggests similarly important reflections (e.g.: $P4_2/mnm(330) = 43.6°2\Theta$, $Im\bar{3}m(011) = 43.13°2\Theta$ or $P4_2/mnm(411) = 47.03°2\Theta$, $P6_3/mmc(101) = 47.36°2\Theta$); however, the reconstruction error provides evidence of an unknown structure.

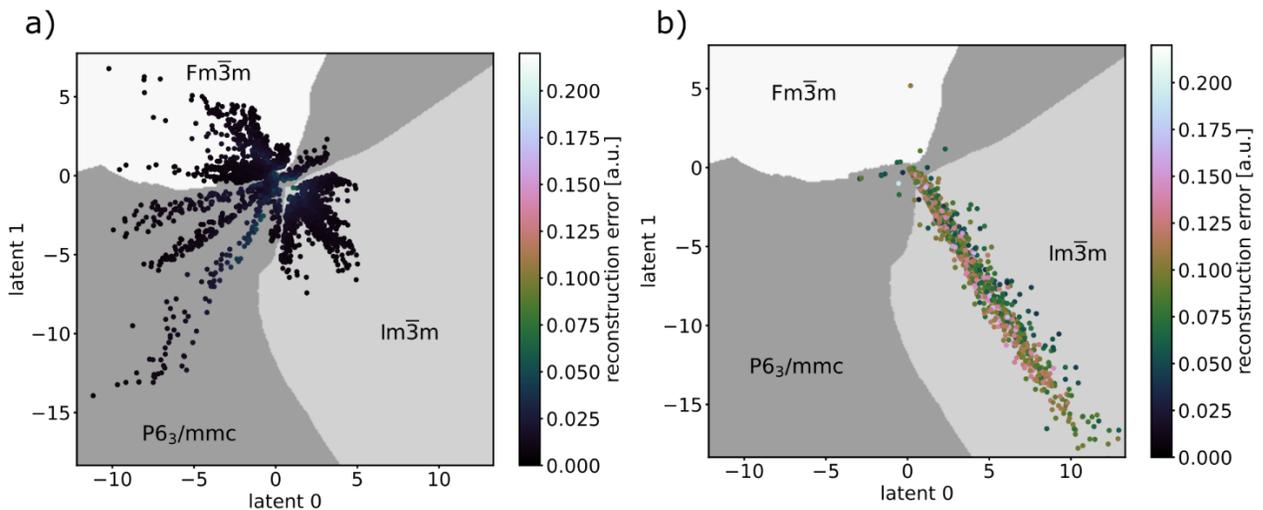

Figure 2: Classification and novelty detection using the reconstruction error. The decision boundaries of a KNN classifier are outlined. a) Latent space representation of known phase color-coded by the reconstruction error. b) Latent space representation of an unknown phase color-coded by the reconstruction error. The unknown phase shows a distinctly higher reconstruction error (average recon. error = 0.09) compared to phases that are recognized by the model (average recon. error = 0.017). The classification test score in this idealized case was 99.16 %.

The reconstruction loss between the input and the decoder can also elicit knowledge of phase mixtures. Phase mixtures could be considered during dataset synthesis by generating binary and ternary mixtures; however, the combinatorial explosion in multinary material systems would drive the dataset size to numbers that cannot be handled efficiently [21]. To test the VAE behavior with phase mixtures, we generate a dataset of binary combinations of the three structures ($Fm\bar{3}m$, $Im\bar{3}m$, $P6_3/mmc$) for 50 different binary compositions between 0 and 100%, and calculate the reconstruction error (Figure 3a). An increase in average reconstruction error of approximately one order of magnitude is observed for mixtures of known phases and a pure unknown phase. Additionally, the distributions show a larger spread. When considering the reconstruction error as a function of phase fraction, the error is maximized at approximately 50% for all binary mixtures (Figure 3b), indicating a maximum in reconstruction error for XRD patterns that are furthest apart from the training data. The large standard deviation results from cases where the mixture of certain XRD patterns with preferred orientation shows similarity to a pure phase. These properties show that unlike contemporary classification models, a VAE has an indication of when it encounters something unfamiliar. The reconstruction error can be used for decision making by alerting a scientist to review the classification results of a corollary classification model[21–23], guiding an acquisition function for curiosity or exploration driven experiments, or to select an appropriate de-mixing model for multi-phase classification[24].

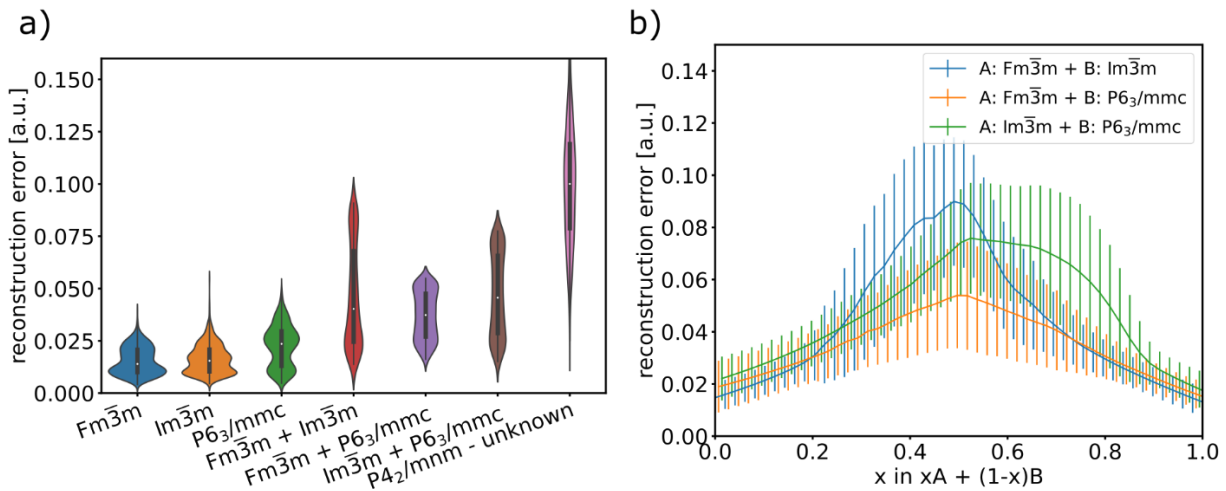

*Figure 3: VAE reconstruction error. a) Violin plot showing the statistics of the reconstruction error for pure phases, two-phase mixtures and an unknown pure structure. The median of phase mixtures is significantly higher than for pure phases. The median of the unknown phase shows an one order of magnitude increase with respect to pure phases. Pure phases show a multimodal distribution, phase mixtures a bimodal distribution and the unknown phase a normal distribution. A broader distribution is observed for phase mixtures and the unknown phase. b) Reconstruction error versus mixing ratio of binary mixtures. The reconstruction error shows a maximum at appr. 50% mixture for which the VAE has the highest uncertainty. We suggest that the reconstruction error be used as a metric for uncertainty that indicates XRD patterns that are either mixtures of phases or new phases that are not contained in the training data. The reconstruction error could be used to guide a data-driven acquisition function in the search for single phase regions in large chemical composition spaces and the discovery of new materials.*

We further tested the VAE in an experimental setting where a multitude of candidate phases are possible. An exemplary experimental dataset acquired from a materials library of the thin film system Co-Ni-Cr-Re contains 225 XRD patterns. Nine quasi-ternary composition spreads are distributed on a rectangular grid over a 100 mm substrate (Fig. SI 1). We selected 21 possible structures from the ICSD database and generated a synthetic dataset following the procedures outlined by Maffettone et al.[8]. The dataset contains several phases that show similar crystal structures for different chemical compositions. The chemical composition is a natural constraint in XRD analysis that bounds phase stability. A sensible alternative condition, beyond the scope of this work, is the phase transition temperature. The chemical composition acts as a constraint using a conditional VAE (cVAE)[25]. In this case, the cVAE encodes the concatenation of chemical composition and 1D XRD pattern (SI). Latent variables and chemical composition are concatenated and passed to the decoder. The relation between chemical composition and crystal structure is coded in the cVAEs parameters after training on the selected 21 structure from database entries.

The VAE is now conditioned on the prior of both the XRD pattern and the chemical composition, which implicates that a mismatch between the conditional (chemical composition) and the XRD pattern will respond in an increased reconstruction error. We exemplify this behavior by recreating the above approach with synthetic data, but using a cVAE (Figure SI2-SI3). Following training on the synthetic data, the reconstruction error of the experimental data is evaluated and mapped over the physical coordinates (Figure 4a). We inspect three XRD patterns with different reconstruction errors in Figure 4b. We also demonstrate the veracity of the latent representation by training a Gaussian Naïve Bayes (GNB) classifier on an ensemble of cVAEs to predict space group from latent vector and chemical composition (Figure SI4).

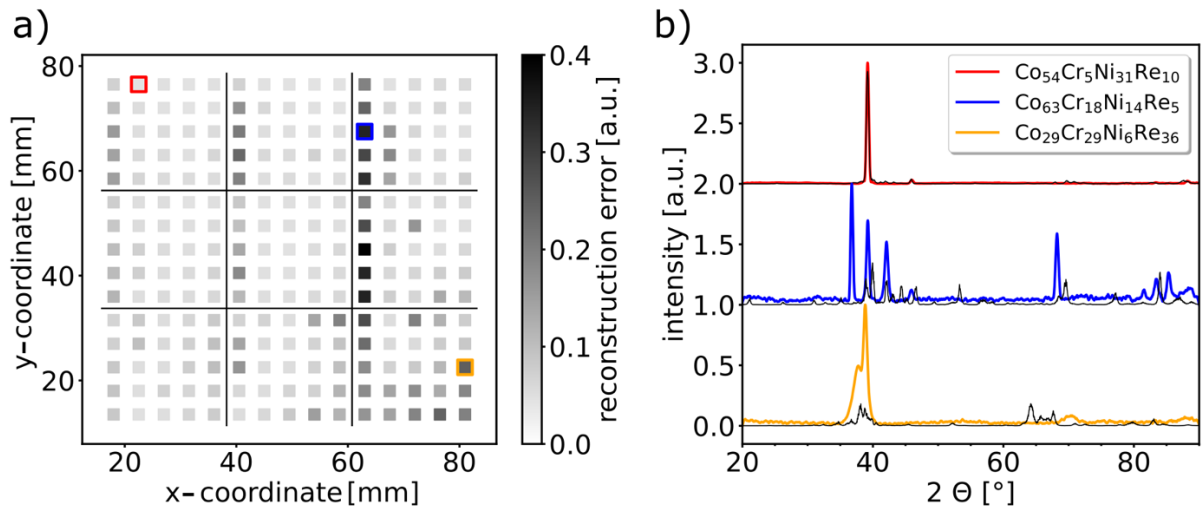

*Figure 4: XRD pattern reconstruction by VAE and reconstruction error as a measure of novelty. a) Materials library map of reconstruction error. Each position marks the measurement area where an XRD pattern was acquired for a unique chemical composition within the library. b) Experimental XRD patterns (color-coded) and VAE reconstructed-XRD patterns (black) of highlighted data points on materials library.*

The sample with the composition $Co_{54}Cr_5Ni_{31}Re_{10}$ shows a highly textured $Fm\bar{3}m$ – Ni structure that was part of the synthetic dataset. Hence, the reconstruction error is minimal and the VAE-reconstructed XRD pattern closely matches the experimental pattern. The sample $Co_{63}Cr_{18}Ni_{14}Re_5$ shows a phase mixture of $Fm\bar{3}m$ – Ni and $P6_3/mmc$ – Co. As phase mixtures are unknown by the model, the reconstruction error is amplified, which informed us that closer inspection of this XRD pattern is required. The sample $Co_{29}Cr_{29}Ni_6Re_{36}$ shows a large peak at 39° 2Θ that matches to $Fm\bar{3}m$ – Ni and a broad side-peak that could be associated with $P4_2/mnm$ according to the GNB prediction.

In summary, we have demonstrated that variational autoencoders provide a powerful toolset to assist in the analysis of XRD datasets that is complementary to other analysis methods. The crystal structure representation provided by the VAE latent vectors highlights intrinsic features of the dataset. This representation can be used for on-the-fly analysis of a dataset distribution across different structures, in which latent space proximity is a structural estimator. The reconstruction error informs the experimenter or adaptive AI driving an autonomous experiment of suspicious data, and assists in detecting unknown phases or phase mixtures. We envision the VAE working cooperatively together in a federation of specialized AI agents. Combining the power of modern classification agents[8,21,24], data reduction agents[26], and the wisdom of ignorance offered by the VAE with adaptive experimental agents[6,27] will significantly increase the veracity and efficacy of high-throughput diffraction measurements.

# Methods

The computed X-ray diffraction datasets used in this study were synthesized using a custom python package based on CCTBX[28]. Three crystallography information files (CIF) from three different spacegroups (ICSD-ids: 108347, 108308, 622438) were chosen. 5000 XRD patterns were simulated for Cu Kα radiation in an angular range from 20 – 90 °2Θ. For each structure the same parameters that are optimized in a Rietveld refinement are varied. The angular range is propagated on 2048 datapoints. The 21-structure dataset for the Co-Ni-Cr-Re system was generated using the same parameters.

The variational autoencoders were written with Python 3.7, using Tensorflow 2.1 and the Keras module. The encoder network consists of an input layer (2048) followed by two dense layers for the encoder (sizes 256 and 128) and a latent space with 2 dimensions (μ and σ). A sampling layer z randomly samples datapoints from a normal distribution (μ = 0 and σ = 1). The output of z is connected to two dense layers of the decoder (sizes 128 and 256) that is connected to an output layer (2048). The loss function is the weighted sum of the binary cross entropy between input XRD and decoded output and the KL-divergence calculated from the latent space distribution and a 2D Gaussian function. The reconstruction loss is weighted by a factor of 2048 to obtain a clear separation of structural features in the latent space. Binary phase mixtures are calculated according to the formula $xXRD_A + (1 − x)XRD_B$ and subsequent normalization of the intensity to 0 and 1.

The conditional VAE used on the experimental dataset comprises the concatenation of the 1D XRD and the normalized chemical compositions as inputs. The latent variable output and the chemical composition are concatenated and provide the input to the decoder.

The Co-Ni-Cr-Re thin film composition spread materials library was deposited in a high vacuum magnetron sputter system (DCA Finland). Deposition was conducted from four 100 mm magnetron sputter cathodes with pure elemental targets in Ar at a pressure of 0.66 Pa and no intentional heating of the substrate. The Ar flow was 60 sccm. Composition gradients were created by using a computer-controlled substrate shutter. The quaternary composition spread type thin film materials library was fabricated by a wedge-type multilayer technique by successive deposition of 200 nanoscale wedge-type elemental layers. A similar approach is described in detail by Salomon et al.[29]. The wedge-type layers were made using a moving shutter, which was set to shield the substrate and was then slowly retracted during deposition. In case of Co and Ni single wedges with length of 67.5 mm were deposited with a rotation of 90° in between. In case of the Cr and Re wedge triplets consisting of three wedges with a length of 22.5 mm were deposited again with a rotation of 90° in between. This the nominal thickness of the wedges varies from 0 to ~13 nm in case of Co, Cr and Ni and from 0 to ~3 in case of Re. After deposition, the materials library was annealed in Ar at $1.1*10^5$ Pa. XRD measurements were done in a Bruker D8 Discover with a 2D detector (Vantec 500) with Cu Kα radiation. Four 2D-frames were measured to cover a 2θ range from 20° to 100°. The frames were merged and integrated into 1D-intensity diffractograms. The background was subtracted by fitting a 4-degree polynomial function.

**Data availability**

The exact datasets used in this study are available from the authors on reasonable request, with the available code used to generate the training data below.

**Code availability**

The XCA package was used to generate training datasets and is available at github.com/maffettone/xca. This package is actively being extended to support VAE models.

Additional code used in this study is available from the corresponding author on reasonable request.

**Contributions**

LB and PM conceptualized the study and wrote the main parts of the manuscript. LB performed the data experiments. DN was responsible for the synthesis and characterization of the experimental Co-Ni-Cr-Re materials library. All authors contributed in the discussion of the results and participated in writing the manuscript.

**Acknowledgements**

This study was funded by the German Research Foundation (DFG) as part of Collaborative Research Centers SFB-TR 87 and SFB-TR 103 and the BNL Laboratory Directed Research and Development (LDRD) project 20-032 "Accelerating materials discovery with total scattering via machine learning". The center for interface dominated high-performance materials (ZGH, Ruhr-Universität Bochum, Bochum, Germany) is acknowledged for X-ray diffraction experiments.